\documentstyle[12pt,epsfig]{article} \input{psfig} 

\title{General relativity and quintessence explain the Pioneer anomaly}

\author{J.P.  Mbelek \\ Service d'Astrophysique, C.E.  Saclay \\ F-91191
Gif-sur-Yvette Cedex, France}

\begin{document} \maketitle \baselineskip=8mm

\begin{abstract} The anomalous time depending blueshift, the so-called "Pioneer
anomaly", that was detected in the radio-metric data from Pioneer 10/11, Ulysses
and Galileo spacecraft may not result from a real change of velocity.  Rather,
the Pioneer anomaly may be understood within the framework of general relativity
as a time depending gravitational frequency shift accounting for the time
dependence of the density of the dark energy when the latter is identified with
quintessence.  Thus, instead of being in conflict with Einstein equivalence
principle, the main Pioneer anomaly appears merely as a new validation of
general relativity in the weak field and low velocity limit.  \end{abstract}

\section{Introduction} Since 1998, Anderson {\sl et al.}  have continuously
reported an anomalous Doppler shift derived from a twenty years study of
radio-metric data from Pioneer 10/11, Ulysses and Galileo
spacecraft~\cite{Andersona}.  The observed effect mimics a constant acceleration
acting on the spacecraft with magnitude $a_{P} = (8.74 \pm 1.33) \times
10^{-8}$~cm~s$^{-2}$, directed toward the sun to within the accuracy of the
Pioneers' antennas and a steady frequency drift $\frac{d\Delta\nu}{dt} \simeq 6
\times 10^{-9}$~Hz/s which equates to a "clock acceleration"
$\frac{d(\Delta\nu/\nu)}{dt} = \frac{a_{P}}{c}$.  An independant analysis of
radio Doppler tracking data from the Pioneer 10 spacecraft for the time period
$1987$ - $1994$ confirms the previous observations~\cite{Markwardt}.  Besides,
it has been noted that the magnitude of $a_{P}$ compares nicely to the Modified
Newtonian Dynamics (MOND) acceleration constant $a_{0} \sim
cH_{0}$~\cite{Milgroma}, where $H_{0}$ is the Hubble parameter at present
cosmological epoch.  Actually, the reported anomaly cannot really be related to
MOND.  Indeed, the gravitational pulling of the sun up to $100$ AU still yields
an acceleration greater than $a_{0}$ by at least three orders of magnitude,
equating $a_{0}$ only at about $3000$ AU.  Hence, Newtonian dynamics up to GR
corrections should apply to spacecraft.  The same argument was put foreward by
M.  Milgrom~\cite{Milgromb} to reject the claim that MOND fails in the
laboratory~\cite{Abramovici}.  In a further study, Anderson {\sl et al.}  found
that the small difference of magnitude of $a_{P}$ for Pioneer 10 and Pioneer 11
is related to their difference of spin-rate history~\cite{Andersonb}.  Removing
the spin-rate change contribution, yields an apparent anomalous acceleration
$a_{P} = (7.84 \pm 0.01) \times 10^{-8}$~cm~s$^{-2}$ of the same amount with a
great accuracy during a long time interval for both Pioneer 10/11.  Since this
has found a conventional explanation~\cite{paper1}, it points out in favor of an
external origin for the main Pioneer anomaly at the expense of any possible
internal cause ({\sl e.  g.}, heat systematics).  As yet, this also excludes any
significant drag force acting on the spacecraft ({\sl e.  g.}, the mirror world
explanation~\cite{Foot}) as an explanation of the anomaly.  Indeed, the
velocities of the Pioneer 10 and 11 relative to the sun differ by about $5 \%$
in the time interval of interest.  Moreover, Whitmire and Matese~\cite{Whitmire}
have carried out a statistical analysis by comparing the mean original energy
obtained from the Comet Catalogue of Marsden and Williams~\cite{Marsden}, with
and without corrections due to an hypothetical Pioneer anomalous acceleration,
to that predicted by galactic tidal theory.  They conclude that the implied
higher binding energy is incompatible with the established evidence that the
galactic tide is dominant in making Oort cloud comets observable.  The whole
work of these authors is based on the major assumption that the Pioneer anomaly
reveals a real acceleration.  As they emphasized, systematic outgassing would
not hide the effects of such anomalous acceleration if real.  Therefore, the
alternative that the Pioneer anomaly does not result from a real change in
velocity deserves to be investigated.  Indeed, a direct interpretation of the
observational data from the spacecraft implies merely a time dependent frequency
blueshift of the photons.  However, any true Doppler shift would involve an
accompanying acceleration.  Hence, given our knowledge of photons frequency
shift, the only other relativistic effect that can be confused, at the solar
system scale, with a real Doppler shift is the gravitational frequency shift.
In the weak field and low velocity limit, this should involve a time dependent
gravitational potential instead of a spatial dependent one, in order to avoid an
induced anomalous acceleration for test bodies.  Of course, the origin of a time
dependent gravitional potential would not be relevant if not justified on the
physical ground.  Actually, although unfamiliar, we show that a time dependent
gravitational potential follows from our present knowledge of the matter-energy
content of the universe.  Indeed, quintessence is one possible form of dark
energy that explains the recent discovery of the accelerating universe in term
of a scalar field whose energy density is time dependent and dominant over that
of the ordinary matter at present cosmological epoch.  Now, a time dependent
matter-energy density will naturally generate a time dependent potential.  Given
that the energy density of the quintessence is on average the same everywhere
depending only on time, this should holds even at small scale, not only at the
cosmological level.  The idea of a time dependent gravitational potential have
been revived recently by A.  F.  Ra\~nada~\cite{Ranada} and K.
Trencevski~\cite{Trencevski}.  However, both attempts introduce only
phenomelogically the time dependent potential ({\sl e.  g.}  by choosing its
sign by convenience and trying to match exactly $a_{P}$ with $c H_{0}$).  As a
common feature, none of both approaches refer to a field equation to derive the
time dependent potential.  Above all, none rely intirely on GR.  Besides, we
note that the link between the time dependent potential and dark energy (whether
this pertains to a true cosmological constant or quintessence is not expressed
at all) is either put by hand~\cite{Ranada} or just invoked by
words~\cite{Trencevski} by the authors.  The aim of this paper is to demonstrate
that the Pioneer anomaly can find a natural explanation within the framework of
general relativity (GR) when the gravitational contribution of quintessence is
properly accounted for.  Then, the Pioneer anomaly interprets as a gravitational
blueshift in excess induced locally by the energy density of quintessence which
provides a time dependent contribution to the gravitational potential.
Furthermore, because of its weakness, the Pioneer anomaly seen as a time
dependent gravitational potential actually entails no contradiction with respect
to the four classical tests of GR.  Also, it is worth noticing that the analysis
that is carried out in the following involves the Hubble constant only through
the time dependence and non locality of the energy density of the quintessence.
In any case, the matter content of gravitational bound systems like galaxies or
the solar system are almost decoupled from the general expansion of the universe
whose effect is quite negligible on the stars or planetary orbits
(see~\cite{Misner} and~\cite{Rindlera}, section 16.1 H).  Indeed, the exact
mathematical treatments of gravitational bound systems in an expanding universe
that are based on GR solely (with or without a true cosmological constant)
always lead to conclude to a negligible effect out of an ensemble of dust
particles~\cite{Noerdlinger, Bonnor, JLAnderson} or a star~\cite{Gilbert}.
Thus, although the net effect is not exactly zero, it cannot account for the
Pioneer anomaly in the framework of GR (see {\sl e.  g.},~\cite{Licht}) without
reconsidering the matter-energy content of the universe.

\section{Time varying potential from quintessence} Let us consider the dark
energy as quintessence, that is a neutral scalar field coupled to ordinary
matter only through its gravitational influence.  We show that this involves a
time varying contribution to the gravitational potential even for
gravitationally bound systems.  As one knows, unlike the ordinary matter
(luminous or dark matter) whose mass density tends to decrease continuously
driven by the expansion of the universe or clumps locally under the
gravitational pulling, the energy density of the dark energy remains constant on
average everywhere at any given cosmic time (quintessence models) if not a
universal constant (a property of spacetime itself) in the case of a true
cosmological constant.  For that reason, we will not assume a static metric for
the solar system hereafter.  Instead, in addition to the point like mass of the
sun and the other heavenly bodies (whose effects as usual are seen as
perturbations), we will also deal with the gravitational effect of the uniform
mass density, $\rho_{Q}$, of the dark energy which behaves like a cosmological
"constant" $\Lambda = 8\pi G\rho_{Q}/c^{2}$.  By choosing the coordinates such
that the metric tensor be diagonal, the metric still expresses in the canonical
form \begin{equation} \label{metric} ds^2 = ( \,1 \,+ \,2 \,\frac{V}{c^{2}})
\,c^{2} dt^{2} \,- \, (\,1 \,+ \,2 \,\frac{V}{c^{2}} \,)^{-1} \,dr^{2} \,-
\,r^{2} \,( \,d\theta^{2} \,+ \, \sin^{2}{\theta} d\varphi^{2} \,),
\end{equation} but with the potential $V$ now depending ton both spatial and
time coordinates, where $x^{0} = c t$, $x^{1} = r$ is the radius from the sun,
$x^{2} = \theta$ is the polar angle and $x^{3} = \varphi$ is the azimutal angle.
The function $V$ is derived from Einstein equations $R_{\mu\nu} - \frac{1}{2}
\,( \,R \,+ \,2\Lambda \,) \,g_{\mu\nu} = \frac{8\pi G}{c^{4}} T_{\mu\nu}$, in
the limit of the weak and slowly varying gravitational field.  We proceed as
usual by setting $g_{\mu\nu} = {\eta}_{\mu\nu} + h_{\mu\nu}$, where all the
$h_{\mu\nu}$'s are much less than unity.  Thus, the Einstein equations reduce to
$R_{00} = \frac{4\pi G}{c^{2}} (\rho_{matter} + \rho_{Q} + 3
\frac{P_{Q}}{c^{2}})$ in the weak field and low velocity limit.  Expanding
$g_{00}$ and $g_{11}$ in the first order in $V/c^{2}$ and $R_{00}$ in the first
order of the derivatives of the $h_{\mu\nu}$'s (see {\sl e.
g.},~\cite{Rindlera}, section $15.1$ and~\cite{Kenyon}) but without discarding
as usual the time derivatives, one gets $R_{00} = \frac{1}{2}
\,{\eta}^{\alpha\beta} \left( \,h_{\beta 0,0\alpha} - h_{00,\beta\alpha} +
h_{0\alpha,\beta 0} - h_{\beta\alpha,00} \,\right) = \frac{1}{c^{2}} \,(
\,\frac{1}{c^{2}} \,\frac{\partial^{2} V}{\partial t^{2}} \,+ \,\Delta V \,)$.
Hence, the equation to solve reads \begin{equation} \label{eq 1} \frac{1}{c^{2}}
\,\frac{\partial^{2} V}{\partial t^{2}} \,+ \,\Delta V = 4\pi G \left(
\,\rho_{matter}(\vec{r}) \,+ \,\rho_{Q} \,+ \,3\,\frac{P_{Q}}{c^{2}} \,\right),
\end{equation} where the equation of state of the quintessence is given by
$P_{Q} = w_{Q} \,\rho_{Q} c^{2}$ with equation of state parameter $w_{Q} \simeq
\,- \,1$.  Usually, one solves (\ref{eq 1}) by assuming a static potential.
This yields (up to an integration constant) \begin{equation} \label{purely
static sol} V = V_{N} \,- \,\frac{1}{6} \,\Lambda \,c^{2} \,r^{2},
\end{equation} where the Newtonian potential $V_{N} = V_{N}(\vec{r})$ is a
solution of Gauss equation \begin{equation} \label{eq Gauss} \Delta V_{N} = 4\pi
G \rho_{matter} \end{equation} and the quadratic extrapotential term is related
to the dark energy density.  However, because of the fine-tuning that would be
required both for the cancellation of the vacuum energy (quantum field theory
problem) and the coincidence problem (why the density of the dark energy and
that of the ordinary matter are of the same order just today, whereas both rates
of evolution are quite different), quintessence models are prevailing in the
literature at the expense of a true cosmological constant~\cite{Perlmuttera}.  A
quintessence model involves a dynamical cosmological constant driven by an
almost homogeneous scalar field (spatial fluctuations cancel out on
average)~\cite{Steinhardt}, $Q$, with a suitable self-interaction potential
(most often chosen as an inverse power law of $Q$).  This means that in some
sens, though not exactly, $\Lambda$ is defined as a function of time in such
models.  So, the previous solution (\ref{purely static sol}) which is purely
static does not hold any more.  Thus, while in the case of a true positive
cosmological constant the vacuum solution yields a de Sitter spacetime,
spacetime is rather asymptotically Minkowskian in the case of quintessence whose
energy density decreases monotonously toward zero with respect to time.  Since
$\rho_{Q} = \rho_{Q}(t)$ is non zero everywhere, Birkhoff's theorem cannot
rigorously be invoked to infer a static spacetime.  Let us emphasize that the
generalization of Birkhoff's theorem to the case with a $\Lambda$-term
(cosmological constant) is restricted to the case when the latter is indeed a
true universal constant~\cite{Rindlerb}.  Besides, it is clear that making a
gauge transformation in order to cancel out any time dependent potential term
will not be hepful, since this is associated with a change of reference frame.
Now, we need to study the motion of spacecraft in the same reference frame as
for the planets.  So, let us look for a solution of the form $V = V_{N}(\vec{r})
+ V_{Q}(t)$.  Equation (\ref{eq 1}) then splits into equation (\ref{eq Gauss})
and the following \begin{equation} \label{eq 2} \frac{d^{2}V_{Q}(t)}{dt^{2}} =
\,- \,8\pi G \rho_{Q}(t) c^{2}.  \end{equation} At present cosmological epoch,
$\rho_{Q}(t)$ is a slowly varying function as compared to the time scale that
concerns the Pioneer 10/11 spacecraft.  Therefore, the solution of equation
(\ref{eq 2}) reads in the first approximation \begin{equation} \label{literal
sol eq 2} V_{Q}(t) = \,- \,\frac{3}{2} \,\Omega_{Q} H_{0}^{2} c^{2} (t -
t_{0})^{2} \,- \,3\Omega_{Q} H_{0}^{2} c^{2} (t_{0} - t_{i}) (t - t_{0}) \,+
\,V_{Q}(t_{0}), \end{equation} where $t_{i}$ is the cosmic time at which
$V_{Q}(t)$ passes by its maximum value and $\Omega_{Q} = \Lambda(t_{0})
c^{2}/3H_{0}^{2}$ denotes the density parameter associated with the quintessence
at present cosmological epoch.  Let us point out that one passes from the
non-Newtonian extra potential of relation (\ref{purely static sol}) to relation
(\ref{literal sol eq 2}) by making the substitution $r \rightarrow c (t -
t_{i})$ in (\ref{purely static sol}) then multiplying the whole result by $3$
(because of the isotropy of the $3$-space).  As regards the phenomenology, it
seems reasonable to assume that the linear term of eq(\ref{literal sol eq 2}) is
of the order of the quadratic one early in the past but finally dominates over
it at present epoch.  These conditions are satisfied for $t_{i} \simeq t_{0}/2$
and imply $V_{Q}(t_{0}) \simeq V_{Q}(0)$.  Present cosmological observations
yield $\Omega_{Q} \simeq 0.7$ and $H_{0} t_{0} \simeq 0.96^{+0.09}_{-0.06}$ for
a flat universe~\cite{Perlmutterb}.  Hence, relation (\ref{literal sol eq 2})
yields \begin{equation} \label{sol eq 2} \frac{V_{Q}(t)}{c^{2}} \simeq \,-
\,H_{0}^{2} (t - t_{0})^{2} \,- \,H_{0} (t - t_{0}) \,+
\,\frac{V_{Q}(t_{0})}{c^{2}}.  \end{equation} According to recent
observations~\cite{Riess}, the cosmic jerk when the expanding universe made the
transition from deceleration to acceleration corresponds to a redshift $z_{j} =
0.46 \pm 0.13$ or equivalently a fractional look-back time $(t_{0} -
t_{j})/t_{0} \simeq 0.4$.  So, the choice of parameter made above is consistent
with the natural expectation that the potential energy $U_{Q} = m\,V_{Q}$ of a
test particle of mass $m$ decreases monotonously past $t_{j}$.

\section{Interpretation of the observational data}

\subsection{The main Pioneer anomaly} Since $g_{00} = 1 \,+ \,2 \,(
\,\frac{V_{N} + V_{Q}}{c^{2}} \,)$, the resulting gravitational frequency shift
of photons is derived from the following relation \begin{equation}
\label{gravitational frequency shift} \frac{\nu_{B}}{\nu_{A}} =
\sqrt{\frac{g_{00}(A)}{g_{00}(B)}} \simeq 1 \,+ \,Z_{ordinary} \,+
\,Z_{anomalous}, \end{equation} where $Z_{ordinary} = \frac{V_{N}(A) \,-
\,V_{N}(B)}{c^{2}}$ represents the familiar gravitational frequency shift
whereas $Z_{anomalous} = \frac{V_{Q}(t_{A}) \,- \,V_{Q}(t_{B})}{c^{2}}$ is the
time varying gravitational frequency shift related to the quintessential dark
energy.  For a small time of flight, $\Delta t = t_{B} - t_{A}$, of the photons
of the communication signals between the spacecraft and the earth ($\Delta t$
negligible as compared to the age of the universe), one gets in the first order
approximation on account of relation (\ref{sol eq 2}) \begin{equation}
\label{anomalous shift 2} Z_{anomalous} \simeq \,- \,\frac{1}{c^{2}}
\,\frac{dV_{Q}}{dt}( \,\frac{t_{A} + t_{B}}{2} \,) \,\Delta t.  \end{equation}
Since $\frac{dV_{Q}}{dt} < 0$, relation (\ref{anomalous shift 2}) implies a
systematic blueshift in excess.  For the time interval of interest for the
Pioneer 10/11 spacecraft, the quadratic term of relation (\ref{sol eq 2}) is
negligibly small with respect to the linear one.  On account of relation
(\ref{anomalous shift 2}) above, this involves an anomalous blueshift in excess
\begin{equation} \label{anomalous shift 3} Z_{anomalous} = \frac{a_{P} \,\Delta
t}{c}, \end{equation} where $a_{P} \simeq H_{0} c$ in accordance with
observations.

The above analysis can be pushed further by taking into account the revolution
and rotation of the earth.  Then, one derives the annual and dayly modulations
following the same reasoning as before.  We argue that these modulations are
further evidences for the dependence of the spacecraft anomalous frequency shift
with respect to time, in particular the time of flight of the photons of the
communication signals.

\subsection{The absence of a true anomalous acceleration} Moreover, the
extrapotential $V_{Q}$ will not effect the motion of the spacecraft or heavenly
bodies being time dependent only.  Indeed, in the weak fields and low velocity
limit, the geodesic equation simplifies to $$ \frac{d^{2}\vec{r}}{dt^{2}} = \,-
\,\frac{1}{2} \,\vec{\nabla} g_{00} $$ \begin{equation} \label{eq of motion 1} =
\,- \,\vec{\nabla} V_{N}, \end{equation} where $\vec{r}$ denotes the position
vector of the test body.

\subsection{The steady frequency drift} According to Anderson {\sl et al.}
(see~\cite{Andersonb}, section V - B and C, p.  19), the "clock acceleration"
$a_{t} = d(\Delta\nu/\nu)/dt$ cannot be attributed to a systematic drift in the
atomic clocks of the DSN (Deep Space Network, the network of ground stations
that are employed to track interplanetary spacecraft).  We show that it follows
from the path equation of photons $dP^{\mu} = \,- \,{\Gamma}^{\mu}_{\alpha\beta}
\,P^{\alpha} \,dx^{\beta}$ for the time component $\mu = 0$, where $P^{0} =
\frac{h\nu}{c}$ and $\vec{P} = \frac{h\nu}{c} \,\vec{u}$ define the $4$-momentum
$P^{\mu}$ of a photon of frequency $\nu$ propagating in the direction of the
unit vector $\vec{u}$.  Since ${\Gamma}^{0}_{00} \simeq \frac{1}{c^{3}}
\,\frac{dV_{Q}}{dt} \simeq \,- \,\frac{a_{P}}{c^{2}}$ in weak field and low
velocity limit, one finds after removing the gravitational frequency shift due
to the Newtonian potential solely \begin{equation} \label{clock acc} a_{t}
\simeq \frac{1}{\nu} \,[ \,\frac{d\nu}{dt} \,- \,( \,\frac{d\nu}{dt}
\,)_{V_{Q}=0} \,] \simeq \frac{a_{P}}{c}, \end{equation} as observed.  Let us
notice that relation (\ref{clock acc}) can be derived in the usual simple manner
from the conservation of energy for the photon to which one assigns a
gravitational mass $m = \frac{h\nu}{c^{2}}$ in accordance with the EP and a
kinetic energy $E_{c} = h \nu$.  Further, the other first order correction
implies by the potential $V_{Q}$ comes from the radial component $\mu = 1$.
This brings an additional contribution to the deflection angle of the photons
near the sun of the order $\Delta \delta \sim a_{t} \,\Delta t$ for a time of
flight $\Delta t$.  Clearly, $\Delta \delta$ is by far less than the classical
result $\delta = 1.75$" by more than $7$ orders of magnitude.

\section{Discussion} The study that we have carried out in this paper does not
pretend to bring the last word on the so-called Pioneer anomaly.  We have just
tried to clarify the subject as to the consistency of such an effect with
general relativity.  Although, the results seem encouraging in this respect, we
still have at least two problems to face.  Namely, whereas the slope of the
observed anomalous frequency shift versus time appears almost constant between
$20$ AU and $80$ AU (see {\sl e.  g.}, FIG.  1 of reference~\cite{Andersona} for
Pioneer 10 in the time period $1987$-$1995$), it drops down to zero below $10$
AU (see FIG.  6 or FIG.  7 of reference~\cite{Andersonb}).  Nevertheless,
Ulysses and Galileo which where moving below or about $5.4$ AU (orbital radius
of Jupiter) also have shown an almost apparent constant acceleration in their
radio Doppler and ranging data but with a high correlation with solar pressure
(respectively $0.888$ and $0.99$).  These issues have never been addressed up to
now by any theoretical study.  Of course, the latter observations cannot fit
within the framework of GR as a result of quintessence since the EP would be
strongly violated.  As a way out, we conjecture that mass loss toward the sun
may potentially explain, at least in part if not all, the vanishing of $a_{P}$
observed on both Pioneer $10/11$ spacecraft below $10$ AU.  Indeed, since the
temperature within the Pioneer 10/11 is almost stabilized to about $300$~K while
decreasing with respect to the heliocentric radius, outgassing out of the
spacecraft may occur (note the large error bars on $a_{P}$ below $10$ AU for all
four spacecraft).  Also given that the mass of the Pioneer has decreased by
almost $7\%$ since the date of launch, a mass loss of about $2$~ppm is still
possible during the following ten years after launch.  Because of the
gravitational pulling of the sun, we expert that the ejected mass will be
directed toward the sun.  For both Pioneer 10/11, one just needs an almost
steady rate of mass loss directed toward the sun $\dot{m} \simeq 0.5$ gm per
year during the ten years following the launch of the spacecraft but quickly
decreasing past this time interval.  As for the issue of the high correlation
with the solar pressure in the case of Galileo and Ulysses, given the proximity
of the sun, it is likely that the capture of dust particles of the solar wind
may provide an answer.  Let us recall that the Pioneer 10/11, Galileo and
Ulysses missions were also dedicated to collect dust particles from the solar
wind, all the four spacecraft being endow with dust detectors
on-board~\cite{dust}.  A more detailed study including precise estimates of the
aforementioned effects will be presented elsewhere.

\section{Conclusion} The study of the Pioneer 10/11 spin-rate change histories
points out in favor of the possibility that the cause of the main Pioneer
anomaly is external to the spacecraft.  However, interpreting the main Pioneer
anomaly as a true acceleration leads to some difficulties as regards its effect
on Long-Period Comet Orbits and the Oort cloud.  Especially, assuming a
resisting medium as cause of the anomaly (with the drag acceleration
proportional to the orbital speed or its square) not only necessitates a hole
inside $10$ or $20$ AU to avoid large effects on short period comets such as
Halley and Encke~\cite{Andersonc} (note also how huge such an anomalous
acceleration would be relative to the earth-moon system, if real about the
orbital radius of the earth) but above all also makes difficult to understand
the analogous anomalous accelerations observed on Ulysses and Galileo data
(orbital radius of both spacecraft less than $5.4$ AU).  The alternative view
that we have adopted then consists to consider that the Pioneer anomaly may not
be the result of a new force term.  Indeed, the main Pioneer anomaly may be well
understood within the framework of GR as a time depending gravitational
frequency shift (with respect to the reference frame used by JPL for Doppler
tracking of spacecraft).  The latter is derived from an extra potential whose
source is the time depending energy density of quintessence like dark energy.
Thus, the Pioneer anomaly may accomodate the EP as expressed by the relation
$a_{t} = a_{P}/c$ between the clock "acceleration", $a_{t}$, and the anomalous
"acceleration", $a_{P}$.  Finally, the Pioneer anomaly seems to be rather a new
validation of GR in the weak field and low velocity limit.  Moreover, the
interpretation of the Pioneer anomaly in the framework of GR favors clearly the
case of quintessential dark energy at the expense of a true cosmological
constant.  This exhibits the potentiality of quintessence to help GR gives rise
to some Machian behavior.  The features emphasized above should make clear the
very difference between our new approach and any other one published as yet on
the subject.

\end{document}